\documentclass[lettersize,journal]{IEEEtran}
\usepackage{amsmath,amsfonts}
\usepackage{algorithmic}
\usepackage{algorithm}
\usepackage{array}
\usepackage[caption=false,font=normalsize,labelfont=sf,textfont=sf]{subfig}
\usepackage{textcomp}
\usepackage{stfloats}
\usepackage{url}
\usepackage{verbatim}
\usepackage{graphicx}
\usepackage{cite}
\begin{document}

\title{Passive Handwriting Tracking via Weak mmWave Communication Signals
}
\author{
Chao Yu, Yan Luo, Renqi Chen and Rui Wang
}
\maketitle

\begin{abstract}
In this letter, a cooperative sensing framework based on millimeter wave (mmWave) communication systems is proposed to detect tiny motions with a millimeter-level resolution. Particularly, the cooperative sensing framework is facilitated with one transmitter and two receivers. There are two radio frequency (RF) chains at each receiver. Hence, the Doppler effect due to the tiny motions can be detected via passive sensing respectively at the receivers, and the velocities of the motions can be estimated by integrating the Doppler frequencies. It is demonstrated that the proposed cooperative sensing system is able to track the handwriting with $ 90 $\% error below $ 6 $ mm. Moreover, the proposed cooperative sensing is robust to the strength of received signal. For example, it works even without the line-of-sight paths from the transmitter to the receivers 
or the sensing target, where the received signal strength is not sufficient for timing synchronization or demodulation.
\end{abstract}


\section{Introduction}
Device-free handwriting trajectory reconstruction is an emerging scenario in the field of human-computer interaction (HCI). The absence of wearable devices provide users a greater level of freedom in usage. Wireless communication systems, because of their wide deployment, have significant potential to enable device-free handwriting tracking. In this paper, the handwriting tracking based on mmWave communication systems will be investigated.

There have been a number of works exploiting the channel state information (CSI) estimated for wireless communications in human motion tracking. For example, the Widar system proposed in\cite{qian2017widar} enabled precise tracking of body motion trajectories at the centimeter level by detecting the Doppler frequencies via CSI, where at least $ 6 $ receive RF chains were requested. The IndoTrack system in \cite{IndoTrack} leveraged the Doppler frequency and Angle-of-Arrival (AoA) information from $ 6 $ receive RF chains to track the trajectory. It was shown that the error margin was within $0.48$ m. Furthermore, the CSI-based methods were also extended to investigate the fine-grained handwriting strokes in WiDraw\cite{widraw}, WiTrace\cite{WiTrace} and CentiTrack\cite{CentiTrack}. Particularly, WiDraw harnessed the AoA of incoming wireless signals at the mobile device to track the hand trajectory, with a median error of $ 5 $ cm. WiTrace extracted the phase of signals reflected off the hand, and calculated the distance of movement based on phase shifts. The estimated trajectory was with a median error of $ 2.09 $ cm. CentiTrack estimated the initial position and motion speed of the target hand via the AoAs and Doppler frequencies obtained from $ 6 $ receive RF chains, such that the error of trajectory tracking was suppressed to $ 1.5 $ cm. All the above works were implemented via the sub-$ 6 $GHz WiFi system. The sensing performance was  constrained by the wavelength of signals. Moreover, the CSI-based sensing methods exhibit high sensitivity to received signal strength. This is because a high-quality estimation of CSI is necessary. As a result, the experiments of handwriting tracking in the above works were conducted close to both the WiFi transmitter and receivers, where there were line-of-sight (LoS) paths between each other. 

Passive sensing is another promising approach for human motion detection with half-duplexing data communication transceivers. For example, it was shown in \cite{PassiveHuman01} that human body movements behind a wall can be tracked by exploiting a WiFi signal in passive sensing. It was further demonstrated in \cite{PassiveHuman02} that human breathing could be detected by WiFi passive sensing. In \cite{PassiveHuman03}, a mmWave-based passive sensing system was developed to distinguish different gestures with high accuracy. A link blockage prediction system for mmWave communication was proposed via estimating the trajectory of blockers in \cite{mmAlert}.

In this paper, we continue to show that passive sensing via mmWave communication signals is able to reconstruct handwriting trajectories with weak received signals, where $ 90 $\% reconstruction errors are below $ 6 $ mm. Particularly, the integrated sensing and communication system is composed of one transmitter and at least two receivers, working on the $ 60 $  GHz band. There are two RF chains at the each receiver to facilitate the passive sensing. The handwriting trajectory can be reconstructed by fusing the Doppler frequencies detected at both receivers. In addition to the accurate reconstruction, it is show that the handwriting tracking is even feasible when there is no LoS path from the transmitter to the receivers or the writing hand. To the best of our knowledge, the handwriting tracking via communication signals in non-line-of-sight (NLoS) scenario has not been demonstrated before.

The remainder of this paper is organized as follows. Section \uppercase\expandafter{\romannumeral2} provides an overview of the system. The detection algorithm for Doppler frequency is presented in Section \uppercase\expandafter{\romannumeral3}. Section \uppercase\expandafter{\romannumeral4} describes the handwriting tracking method. The experimental results and error analysis are presented in Section \uppercase\expandafter{\romannumeral5}, and the conclusion is drawn in Section \uppercase\expandafter{\romannumeral6}.

\section{Architecture of Cooperative Passive Sensing}

\begin{figure}[htbp] 
	\centering  
	\subfloat[LoS]{\includegraphics[width=0.45\columnwidth]{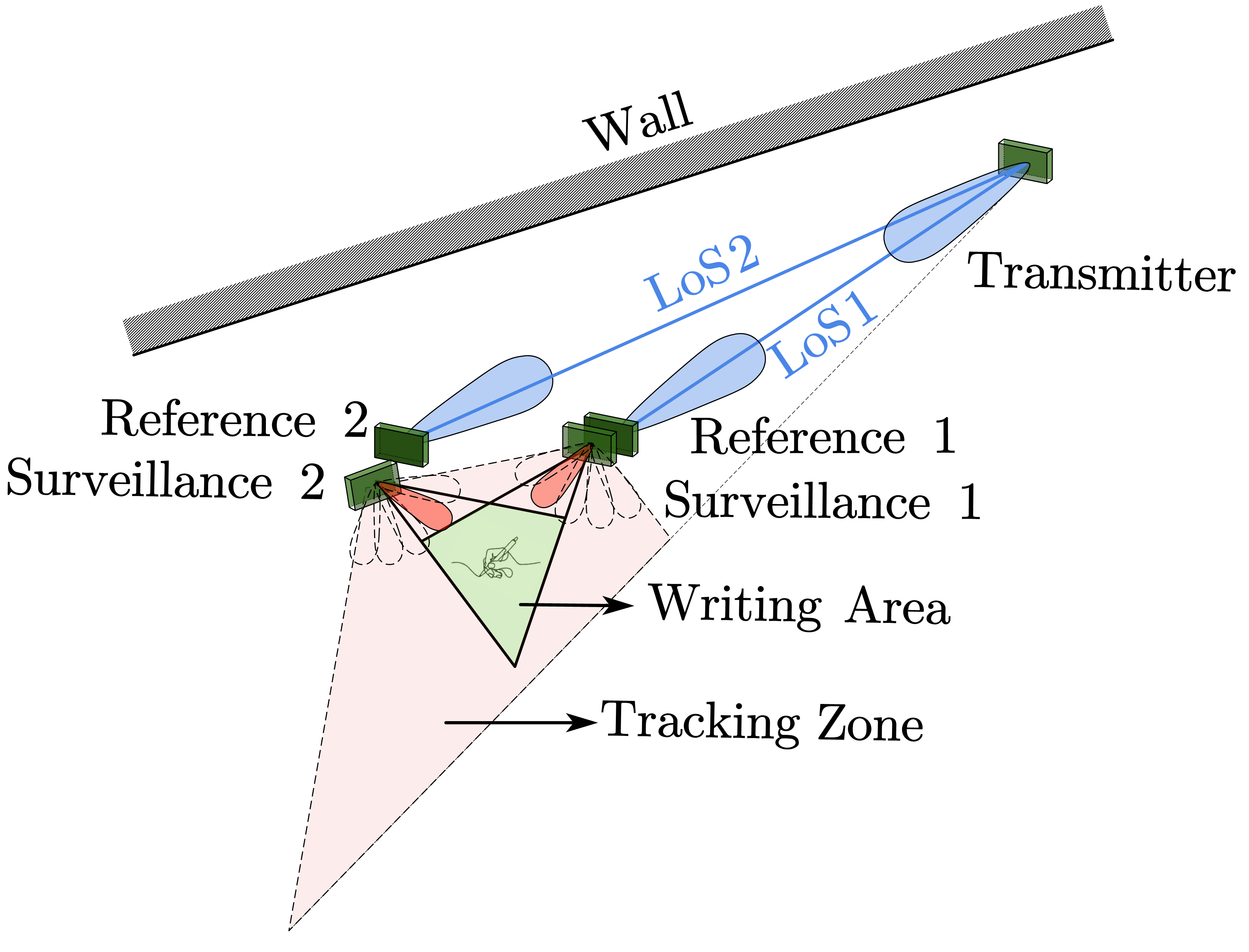}}
	\subfloat[NLoS]{\includegraphics[width=0.45\columnwidth]{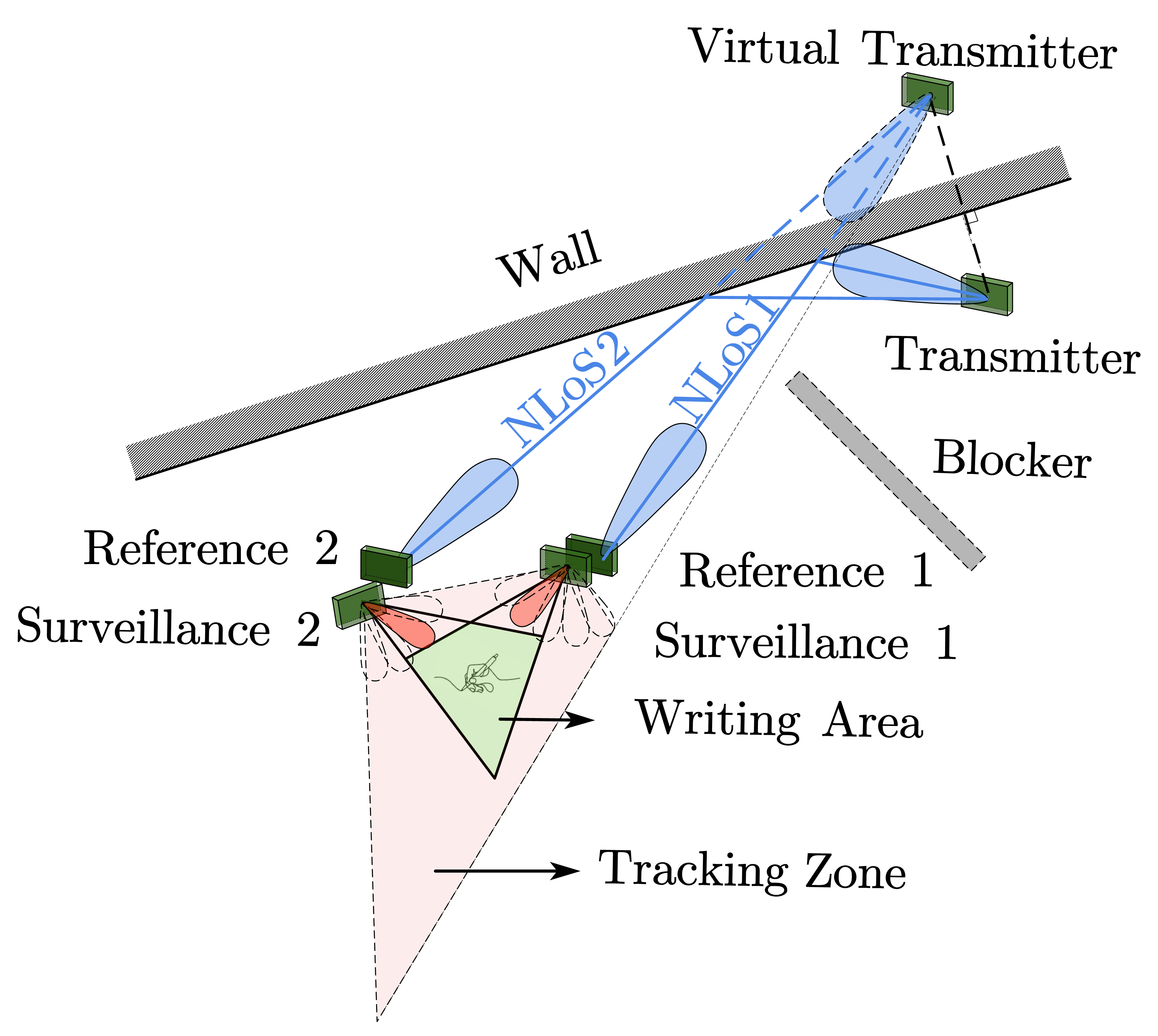}}
	\caption{Example scenarios of cooperative handwriting tracking}
	\label{fig: System overview}
\end{figure}

The proposed system for handwriting tracking consists of one mmWave transmitter and at least two receivers, where each receiver is with two RF chains. The transmitter and receivers can be the base station (BS) and user equipments (UEs) of downlink communications. Thus, two UEs cooperatively detects the handwriting via the downlink signals. The localization techniques have been adopted such that the locations of the transmitter and receivers have been detected. Moreover, time synchronization between the receivers with a maximum error of $ 10 $ ms is assumed for the purpose of data fusion. 

As shown in Fig. \ref{fig: System overview}, the transmitter directs a beam to cover the handwriting tracking zone, including the two receivers and the writing area. The writing pen or finger in the writing area is named as the sensing target in this paper. As a remark note that the path from the transmitter to the handwriting tracking zone can be line-of-sight(LoS) or non-line-of-sight(NLoS).  The two receivers align one of their receive beams with the above LoS or NLoS ray from the transmitter as the ones marked by the blue color in Fig. \ref{fig: System overview}, such that the downlink signal without Doppler effect can be received. The above beams are referred to as the reference beams, and the signal propagation paths are referred to as the reference channels, which provide the baseline signals for handwriting tracking. Meanwhile, each receiver directs the other receive beam to the writing area, which is referred to as the surveillance beam. The signal propagation paths from the transmitter to the receivers scattered off the sensing target in the writing area are referred to as the surveillance channels. Initially, each receiver can 
rotate its surveillance beam, such that the direction of writing area can be obtained by finding the beam direction with the most significant Doppler effect.

The two receivers estimate the Doppler frequencies of their surveillance channels by comparing the received signals of both reference and surveillance channels locally and respectively. The estimated time-varying Doppler frequencies are then synchronized and fused at one of the receivers or the transmitter for handwriting tracking.


\section{Passive Doppler Frequency Detection}

\subsection{Signal Model}\label{sec: signal model}

Let $s(t)$ be the information-bearing signal generated at the transmitter, the received signal via the reference beam at the $i$-th receiver ($i=1,2$) can be written as

\begin{equation}
    y_{r,i}(t)=h_{r,i}s(t-\tau_{r,i})+n_{r,i}(t),\ \  0 \leq t \leq \mathrm{T},
\end{equation}
where $h_{r, i}$ and $\tau_{r, i}$ denote the complex gain and delay of the reference channel respectively, $n_{r, i}(t)$ denotes the noise and interference, $\mathrm{T}$ is the duration of the transmit signal. The interference consists of the signals via the surrounding scattering clusters, e.g., sensing target, the other receiver and etc. Due to the loss of scattering, the interference power is usually much weaker than the signal power in the reference channel. 

Moreover, the received signal of the surveillance beam at the $i$-th receiver ($i=1,2$), denoted as $y_{s, i}(t)$, consists of the  echo signals scattered off the target and surrounding static scattering clusters. Thus, it can be written as
\begin{align}
	\begin{aligned}
		y_{s,i}(t)=
		&h_{s,i}^{\mathrm{tar}}(t)s\left(t-\tau_{s,i}^{\mathrm{tar}}(t)\right)e^{-j2\pi f_{d,i}^{\mathrm{tar}}(t)t}\\
		&+\sum_{l=1}^{L_{i}}h_{s,i}^{l}s(t-\tau_{s,i}^{l})+n_{s,i}(t),\ \  0 \leq t \leq \mathrm{T},
		\label{eq: ys}
	\end{aligned}
\end{align} 
where $h_{s, i}^{\mathrm{tar}}(t)$, $\tau_{s, i}^{\mathrm{tar}}(t)$ and $f_{d, i}^{\mathrm{tar}}(t)$ denote the time-varying complex gain, delay and Doppler frequency of the scattered path off the sensing target respectively, $L_{i}$ is the number of paths from static scattering clusters, $h_{s,i}^{l}$ and $\tau_{s,i}^{l}$ denote the complex gain and delay of the $l$-th static path respectively, and $n_{s,i}(t)$ denotes the noise.

The received signals of both RF chains at the two receivers are sampled at the baseband with a period $\mathrm{T_s}$, which can be expressed as $$y_{r, i}[n] = y_{r, i}(n\mathrm{T_s})\ \mbox{and} \ y_{s, i}[n] = y_{s, i}(n\mathrm{T_s}), $$ where $n=1,2,...,\mathrm{T}/\mathrm{T_s}$ and $i=1,2$. As a remark note that the signal components with zero Doppler frequency in $y_{s, i}$ might interfere the estimation of target Doppler frequency $f_{d, i}^{\mathrm{tar}}(t)$. The least-square-based (LS-based) clutter cancellation elaborated in \cite{tan2005passive} is applied for suppressing the above interference, and the signal of surveillance channel after clutter cancellation is denoted as $\hat{y}_{s, i}[n]$.

\subsection{Doppler Frequency Estimation}\label{sec:Doppler Estimation}
In order to estimate the time-varying Doppler frequencies at both receivers, a sliding-window method based on the  the cross-ambiguity function (CAF) is adopted. Particularly, the Doppler frequencies are estimated every $N_0$ samples, and a window of $N_w$ samples (a.k.a. correlation integration time, CIT) is considered in each estimation. We shall refer to the time instance of the $k$-th Doppler estimation as the $k$-th sensing time instance. Then, the CAF between the reference signal and the surveillance signal of the $i$-th receiver at the $k$-th sensing time instance is defined as
\begin{equation}
    R_{i}(k,f_{d})=\max \limits_{\tau_{i, k}} \sum_{n=kN_0}^{kN_0+N_w-1}\hat{y}_{s,i}[n]y_{r,i}^{*}[n-\tau_{i, k}]e^{-j2\pi f_{d}n\mathrm{T_s}}
	\label{eq: CAF_slide}
\end{equation}
where $(.)^{*}$ is the complex conjugate. It can be observed that a peak value of $R_{i}(k,f_{d})$ will be detected around $f_{d}=f_{d,i}^{\mathrm{tar}}(kN_0\mathrm{T_s})$. Since we focus on the feature extraction of Doppler frequencies in this work, the delay $\tau_{i, k}$ is not considered as a parameter of the CAF.

In order to avoid false alarm, an adaptive-threshold-based method is adopted to detect the Doppler frequency from the CAF. Particularly, a Doppler frequency $f_{d}$ is detected at the $k$-th sensing time instance and the $i$-th receiver, if $$R_{i}(k,f_{d}) \geq \beta_{i}(k,f_{d}),$$
where the threshold $\beta_{i}(k,f_{d})$ can be calculated according to \cite{PassiveHuman02}. Thus, 
\begin{equation}
	\beta_{i}(k,f_{d})
	=
	\frac{\gamma}{2W+1}
	\sum_{p=-W}^{W}
	R_{i}(k,f_{d} + p\Delta f),
	\label{eq:CFAR}
\end{equation}
where $W$ is the half length of training cells, $\gamma>1$ is a scaling factor for detection threshold, and $\Delta f = \frac{1}{\mathrm{T_s}N_w}$ is the resolution of Doppler frequency.  In practice, the detected Doppler frequency may not be unique. We can choose the following strongest component as the detected Doppler frequency:
\begin{alignat}{3}
	\hat{f}_{i}(k)
	&= \ &&\mathop{\arg\max}_{f_{d}}
	R_{i}(k,f_{d})  \nonumber\\
	&s.t. \ &&R_{i}(k,f_{d}) \geq \beta_{i}(k,f_{d}).\nonumber
\end{alignat}

\section{Trajectory Tracking}  

Given the Doppler frequency detection at both receivers, the cooperative handwriting tracking method is elaborated in this section. The tracking can be conduced at either receiver or the transmitter, where the detected Doppler frequencies $\{\hat{f}_{i}(k)|\forall i,k\}$ have been collected. In the following, we first establish the relation between the trajectory of the sensing target and the Doppler frequency, followed by the tracking algorithm.

\subsection{Motion Model}\label{sec: motion model}

\begin{figure}[htbp]
    \centering
    \includegraphics[width=0.8\columnwidth]{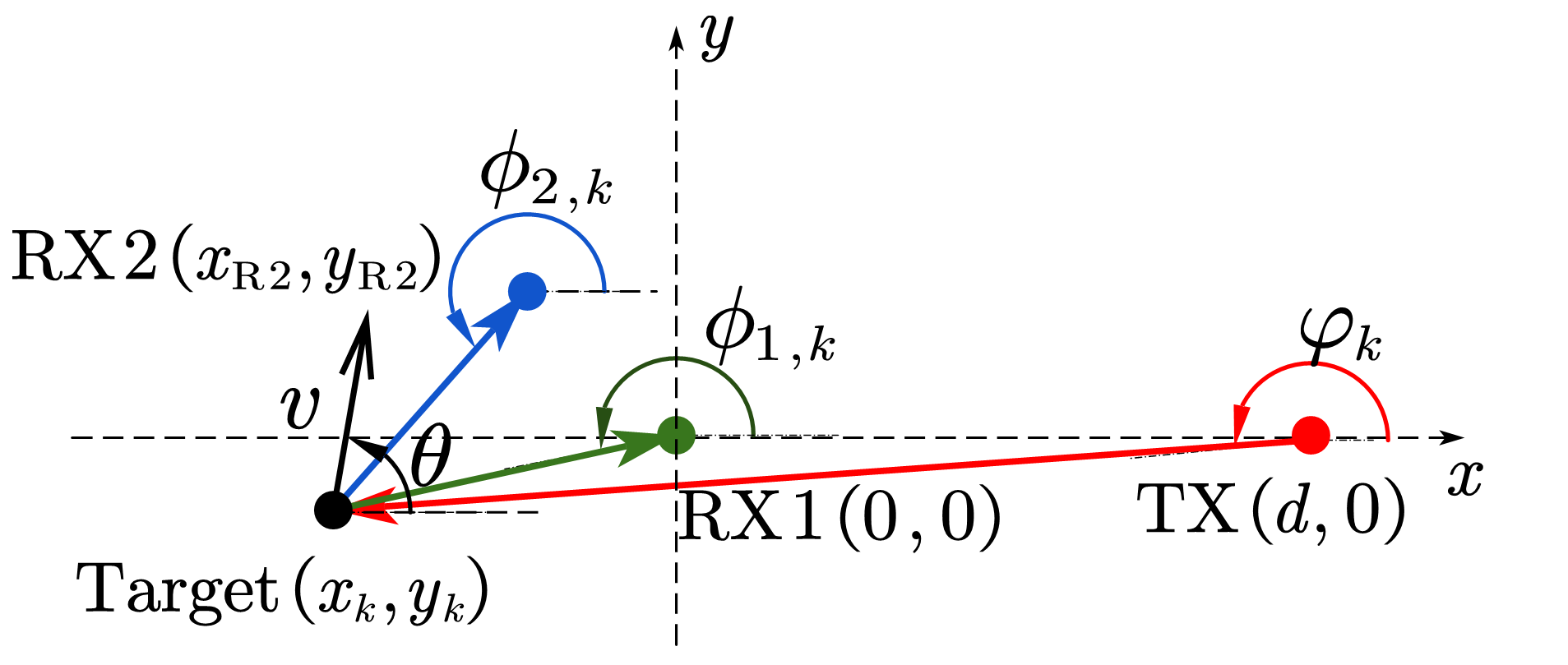}
    \caption{Illustration of AoAs, AoD, and target's mobility.}
    \label{fig: coordinate system}
\end{figure}

As shown in Fig. \ref{fig: coordinate system}, locations of the transmitter and two receivers are denoted as $(d,0)$, $(0,0)$ and $(x_{\mathrm{R2}},y_{\mathrm{R2}})$, respectively. In the NLoS scenario illustrated in Fig. \ref{fig: System overview}, $(d,0)$ represents the location of the virtual transmitter. It is assumed that $d$ and $(x_{\mathrm{R2}},y_{\mathrm{R2}})$ have been estimated via the methods in the existing literature. For example, the directions of the second receiver and the transmitter with respect to the first receiver can be estimated via the MUSIC algorithm \cite{Music}, and the distances can be estimated by multi-tone ranging\cite{Multi_Tone}.

Let $(x_k,y_k)$ be the location of the target at the $k$-th sensing time instance,  $\phi_{1,k}$, $\phi_{2,k}$ and $\varphi_{k}$ be the corresponding AoAs and AoD of the surveillance channel at the two receivers and the transmitter respectively, we have
\begin{align}
	\begin{cases}
		\phi_{1,k}&= \arctan\left(\frac{y_k}{x_k}\right)\\
		\phi_{2,k}&= \arctan\left(\frac{y_k-y_{\mathrm{R2}}}{x_k-x_{\mathrm{R2}}}\right)\\
		\varphi_{k}&= \arctan\left(\frac{y_k}{x_k-d}\right).
		\label{eq: AoA&AoD}
	\end{cases}
\end{align}
Moreover, let $v_k$ and $\theta_k$ be the speed and direction of the target's motion, the Doppler frequencies of the target sensed by the two receivers at the $k$-th sensing time instance, denoted as $f_{1,k}$ and $f_{2,k}$, are given by
\begin{align}
	\begin{cases}
		f_{1,k}=-\frac{2f_c}{c}v_k\cos\left(\theta_k-\frac{\phi_{1,k}+\varphi_{k}}{2}\right)\cos\left(\frac{\phi_{1,k}-\varphi_{k}}{2}\right)\\
		f_{2,k}=-\frac{2f_c}{c}v_k\cos\left(\theta_k-\frac{\phi_{2,k}+\varphi_{k}}{2}\right)\cos\left(\frac{\phi_{2,k}-\varphi_{k}}{2}\right),
		\label{eq: Doppler}
	\end{cases}
\end{align}
where $f_c$ and $c$ denote carrier frequency and light speed, respectively. 

Finally, the location update of the target at the $(k\text{+}1)$-th sensing time instances is given by 
\begin{align}
	\begin{cases}
		x_{k+1}=x_{k}+v_k N_0\mathrm{T_s} \cos\theta_k
		\\
		y_{k+1} = y_{k}+v_k N_0\mathrm{T_s} \sin\theta_k.
		\label{eqn:x_k}
	\end{cases}
\end{align}

\subsection{Handwriting Trajectory Estimation}
Let $\phi_{1,1}$ and $\phi_{2,1}$ be the initial AoAs of the sensing target with respect to the two receivers at the first sensing time instance. They can be estimated by rotating the surveillance beams and finding the beam directions with the strongest Doppler effect. Hence, the initial position of the target, denoted as $(x_1,y_1)$, can be obtained by solving the following equations:
\begin{align}
	\begin{cases}
		\tan(\phi_{1,1}) = \frac{y_1}{x_1}
		\\
		\tan(\phi_{2,1}) = \frac{y_1-y_{\mathrm{R2}}}{x_1-x_{\mathrm{R2}}},
	\end{cases}
\end{align}
As a result, the handwriting trajectory can be tracked iteratively via the following steps, where the iteration index $k$ is initialized with $k=1$.

\textbf{Step 1: velocity estimation.} In \eqref{eq: Doppler}, let $f_{1,k} \approx \hat{f}_{1}(k)$ and $f_{1,k} \approx\hat{f}_{2}(k)$, the speed $v_k$ and direction $\theta_k$ of the target motion can be solved according to \eqref{eq: AoA&AoD} and the knowledge on $(x_k,y_k)$.

\textbf{Step 2: location update.} With the estimation of speed $v_k$ and direction $\theta_k$, the location of the sensing target at the $(k+1)$-th sensing time instance can be obtained from \eqref{eqn:x_k}. Let $k=k+1$, and jump to the Step 1 unless the iteration terminates.

\section{EXPERIMENTS AND DISCUSSION}

\begin{figure}[htbp] 
	\centering  
	\includegraphics[width=1\columnwidth]{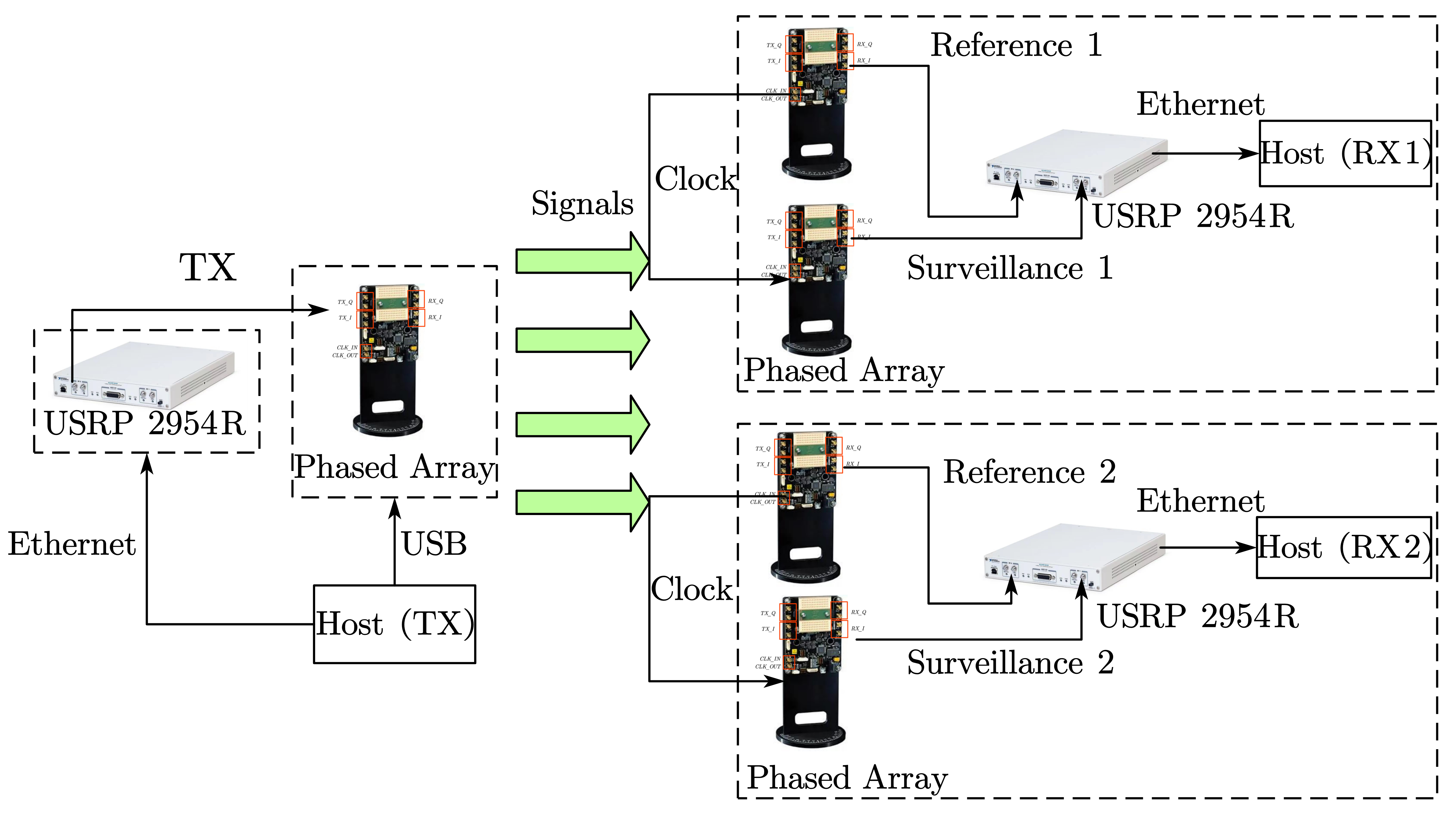}
	\caption{Block diagram of system implementation}
	\label{fig: system}
\end{figure}

In the experiments, the implementation of the proposed system is shown in Fig. \ref{fig: system}. The transmitter is implemented with one NI USRP-2954R connected with one Sivers $60$ GHz phased array. The transmit signal with a bandwidth of $5$ MHz consists of a training sequence and an OFDM-modulated data payload. At each receiver, two Sivers $60$ GHz phased arrays with a common clock ($ 45 $ MHz) are connected to one NI USRP-2954R. The beam widths of both receive phased arrays are $10^\circ$. The phased arrays are configured with $64$ pre-defined beambooks, providing a scanning range of up to $90^\circ$ with an minimum spacing of $1.5^\circ$. At the receive USRPs, the sampling rate $f_s$ is $10$ MHz, the sliding window duration is $N_w \mathrm{T_s} = 0.1$ s, and the interval between two sensing time instances is $N_0 \mathrm{T_s} = 0.01$ s.

\begin{figure}[htbp] 
	\centering
	\subfloat[LoS]{\includegraphics[width=0.49\columnwidth]{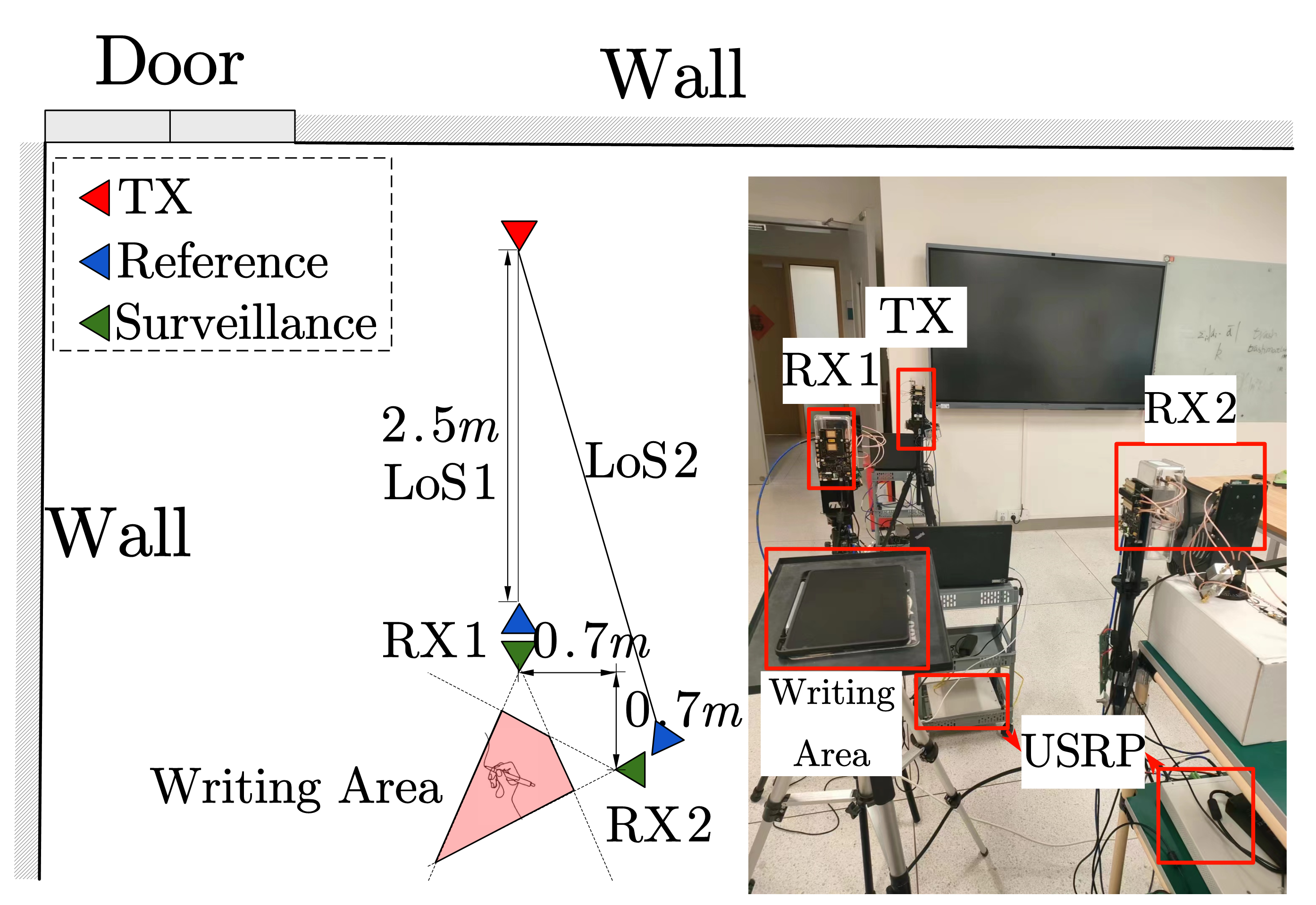}}
	\subfloat[NLoS]{\includegraphics[width=0.49\columnwidth]{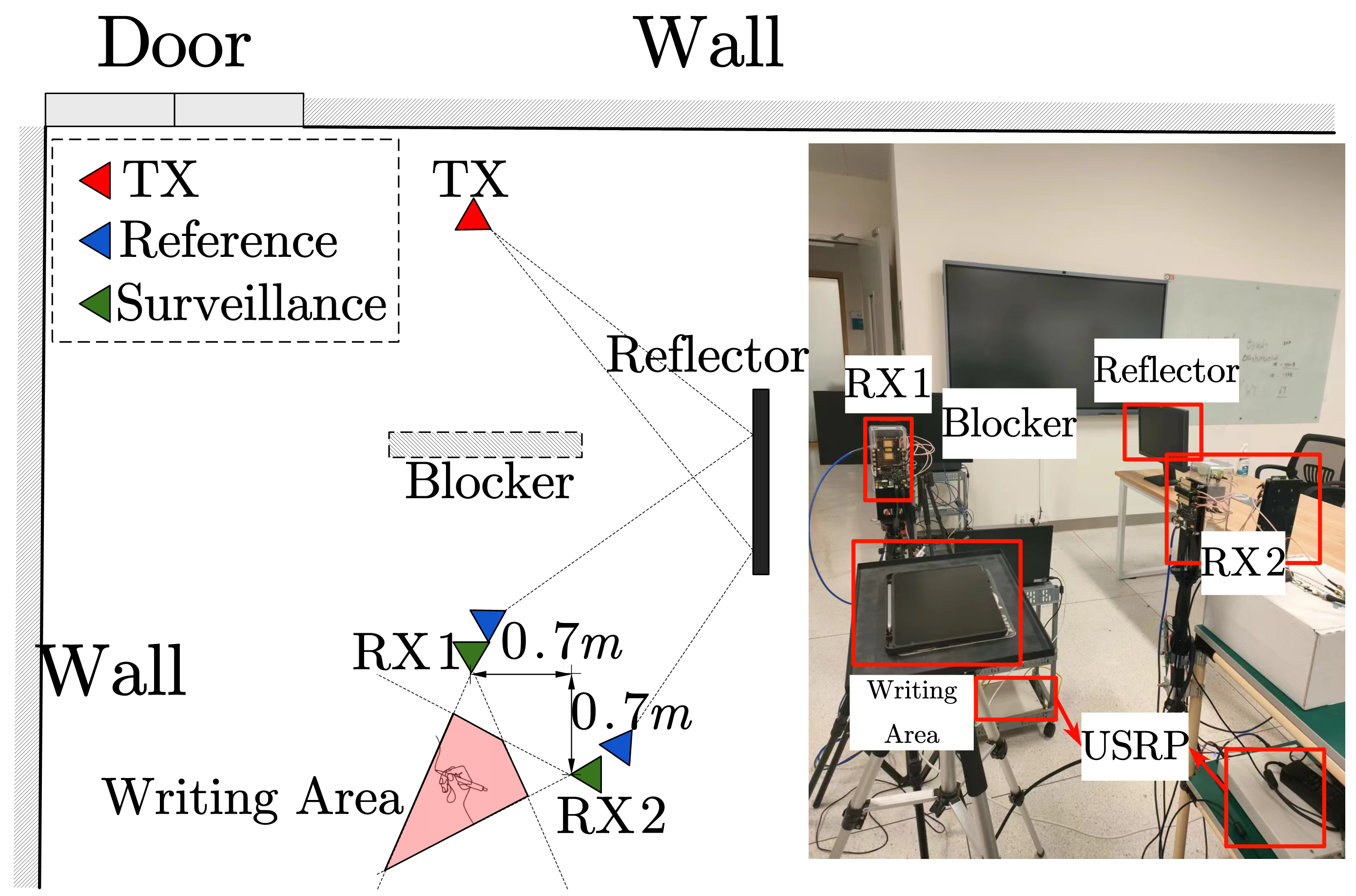}}
	\caption{Experimental scenarios.}
	\label{fig: experiment scenario}
\end{figure}

The experiments are conducted in a laboratory environment with rich scattering clusters, e.g., displays and metallic cabinets. The placement of the two receivers and the writing platform is illustrated in Fig. \ref{fig: experiment scenario}, where one volunteer is writing on a pad such that the ground truth of the writing trajectories can be recorded. Both LoS and NLoS scenarios are considered in the experiments. In the LoS scenario shown in Fig. \ref{fig: experiment scenario} (a), the transmitter is $ 2.5 $ m away from receiver 1, while there is $ 1 $m separation between the two receivers.  In the NLoS scenario shown in Fig. \ref{fig: experiment scenario} (b), the LoS paths from the transmitter to the two receivers and the sensing target are all blocked. Hence, the cooperative sensing relies on the NLoS signal reflected off the display and wall.

\subsection{Tracking Results in LoS Scenario}

\begin{figure}[htbp] 
	\centering  
	\subfloat[RX1]{
		\includegraphics[width=0.45\columnwidth]{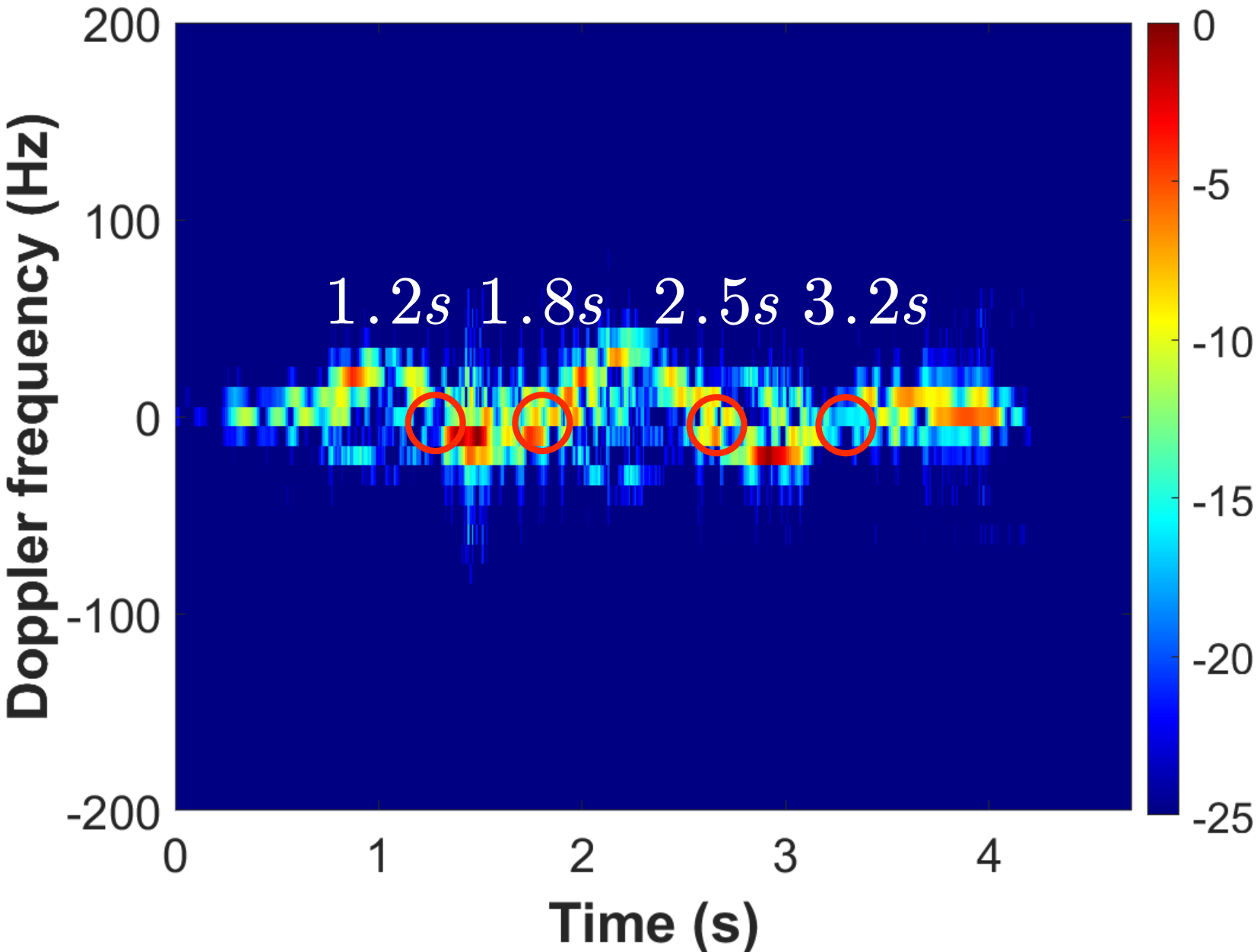}}
	\subfloat[RX2]{
		\includegraphics[width=0.45\columnwidth]{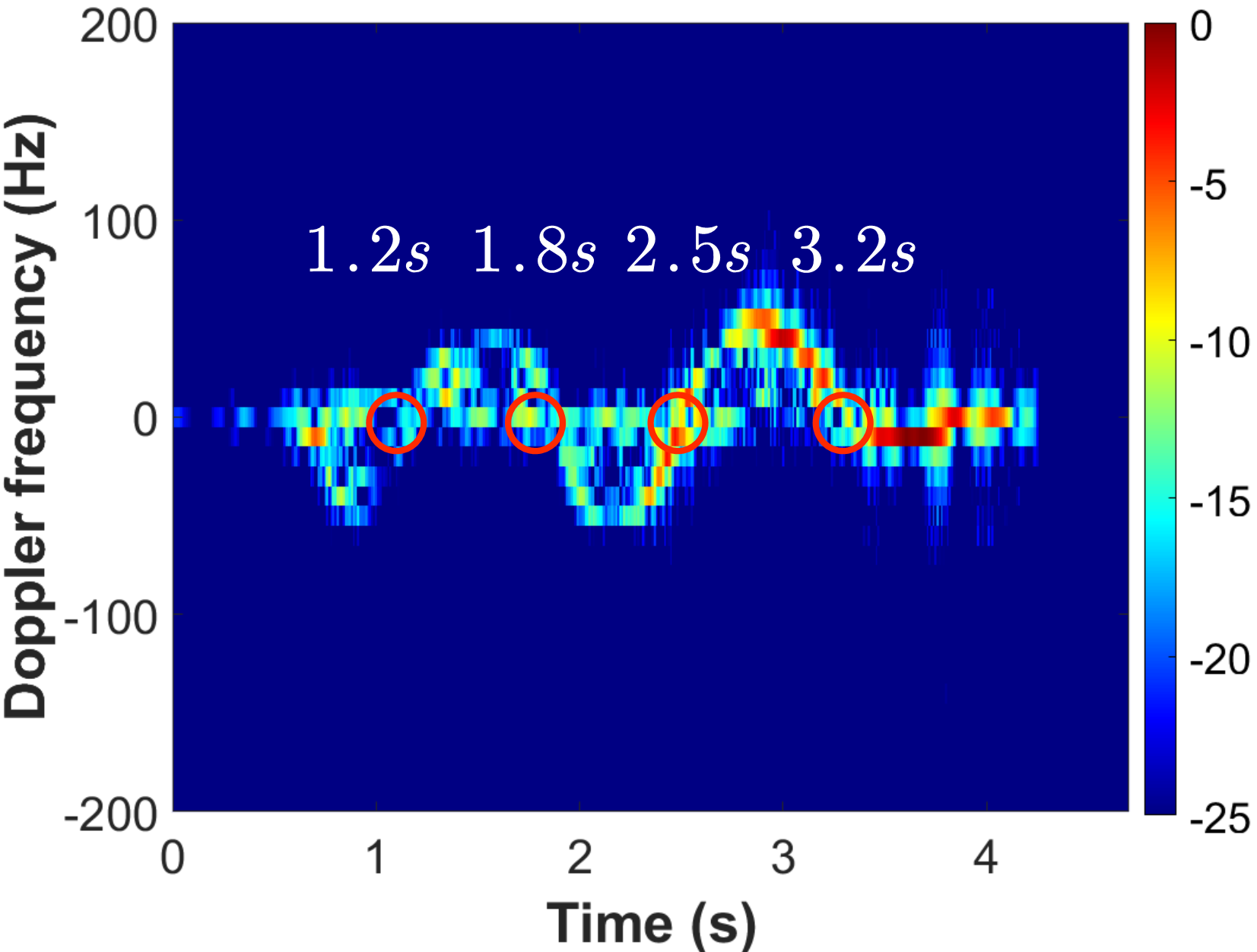}}
	\caption{The measured time-Doppler spectrograms at the two receivers.}
	\label{fig: time-Doppler spectrogram}
\end{figure}

The results of handwriting tracking in the LoS scenario are presented in this part. First, the detected Doppler frequencies versus time at the two receivers are illustrated in Fig. \ref{fig: time-Doppler spectrogram}, where the digit "3" is written. It can be observed that the duration of handwriting is from $ 0.6 $ s to $ 4.4 $ s. The Doppler frequency differs at the two receivers. It reaches a maximum value of $ 50 $ Hz at $ 2.2 $ s at the first receiver; meanwhile, the measured Doppler frequency is $  -50 $ Hz at the second receiver. Moreover, zero Doppler frequency can be observed at both receivers at the time instances $ 1.2 $ s, $ 1.8 $ s, $ 2.5 $ s and $ 3.2 $ s. 
This is because of the temporary stops (turning points) of the writing. 

\begin{figure}[htbp]
    \centering
    \includegraphics[width=0.6\columnwidth]{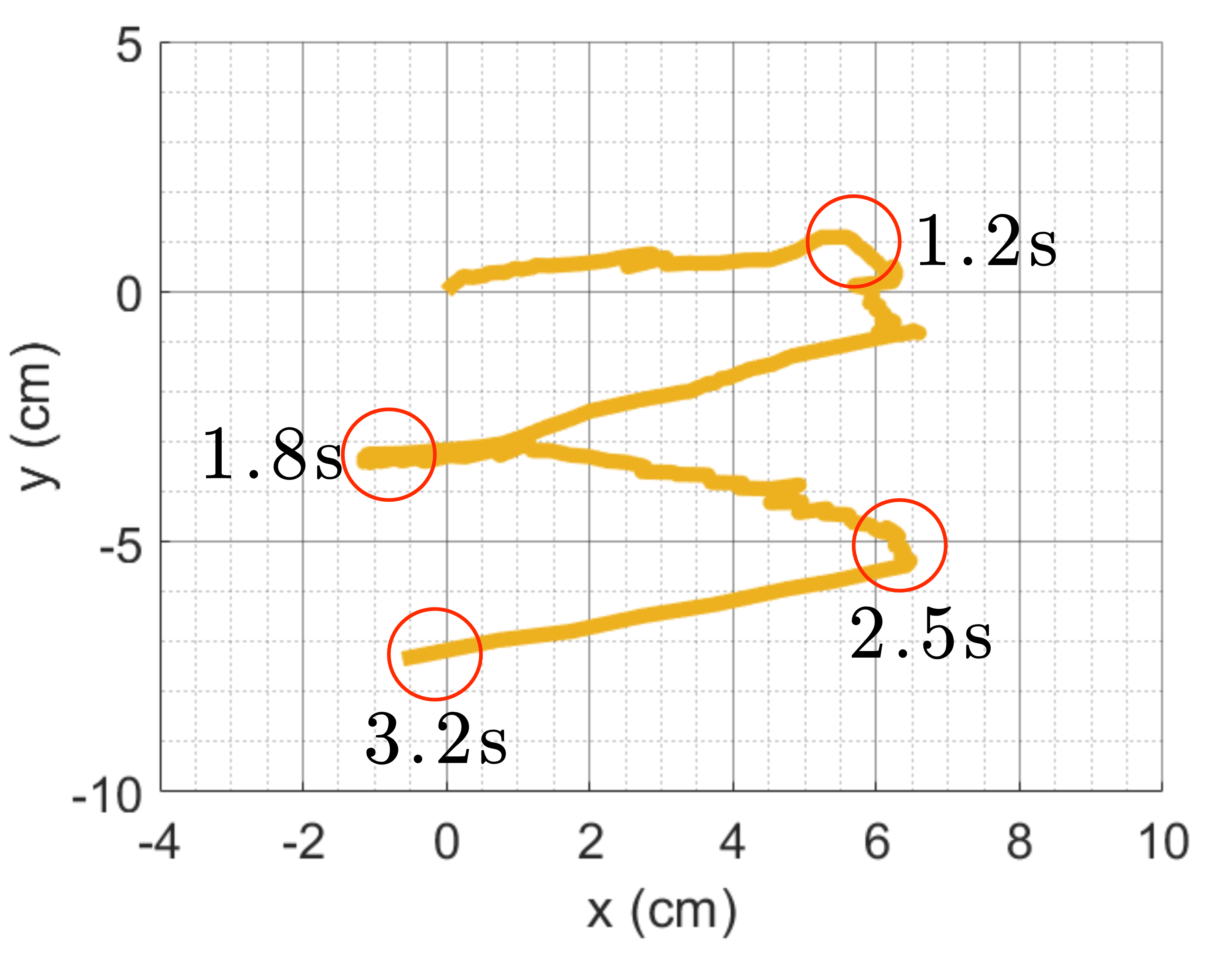}
    \caption{Reconstruction of the handwriting trajectory of the digit 3 in LoS scenario.}
    \label{fig: number 3 reconstruction}
\end{figure}

The reconstructed trajectory is depicted in Fig. \ref{fig: number 3 reconstruction}. It can be observed that the digit can be clearly identified. Moreover, by comparing Fig. \ref{fig: number 3 reconstruction} and \ref{fig: time-Doppler spectrogram}, zero Doppler can be found at the three turning points of the trajectory, as marked in the Fig. \ref{fig: number 3 reconstruction}. 

\subsection{Tracking Results in NLoS Scenario}

\begin{figure}[!t] 
	\centering  
	\subfloat[Digit 3]{
		\includegraphics[width=0.45\columnwidth]{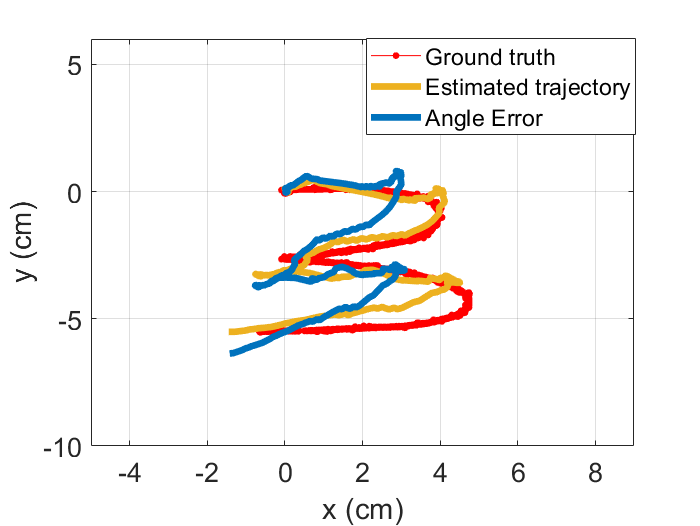}}
	\subfloat[Star]{
		\includegraphics[width=0.45\columnwidth]{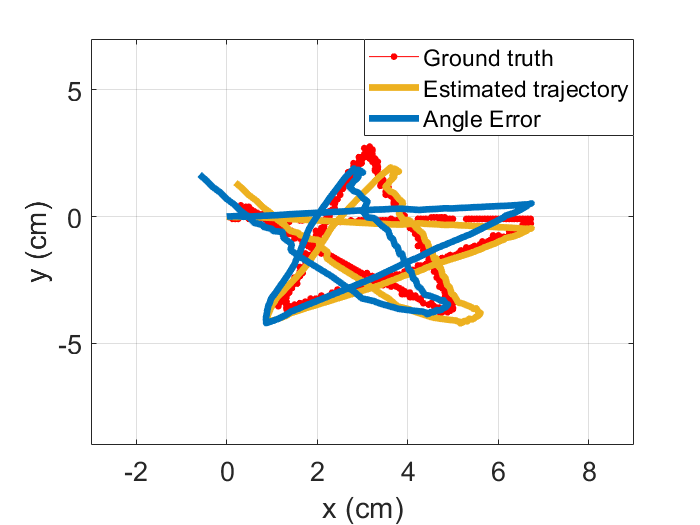}}
	\caption{Handwriting trajectory tracking results in NLoS Scenario.}
	\label{fig: Trajectory(NLoS)}
\end{figure}

The potential of passive sensing in handwriting tracking is further exploited in the NLoS scenario, where the received signal power is significantly degraded due to the loss in scattering and reflection. In this scenario, the volunteer first writes a digit "3", and then draws a more complicated shape, i.e., a star. The reconstructed trajectories and the ground truths are compared in Fig. \ref{fig: Trajectory(NLoS)}. In order to show the reconstruction error, all the trajectories are aligned at the same initial point. Note that reconstructed trajectories depend on the detection of initial point $(x_1,y_1)$, the detection error may raise distortion of the reconstructed trajectories. The trajectory marked by orange uses the correct values of $(x_1,y_1)$, and the trajectory marked by blue uses the wrong values. It can be observed that (1) even with error in the detection of $(x_1,y_1)$, the reconstructed trajectories are still sufficiently clear to identify the digit and the shape; (2) the reconstruction can be very close to the ground truth with the correct values of $(x_1,y_1)$.

\subsection{Robustness Analysis on received SNR}

\begin{figure}[htbp] 
	\centering
	{\includegraphics[width=0.6\columnwidth]{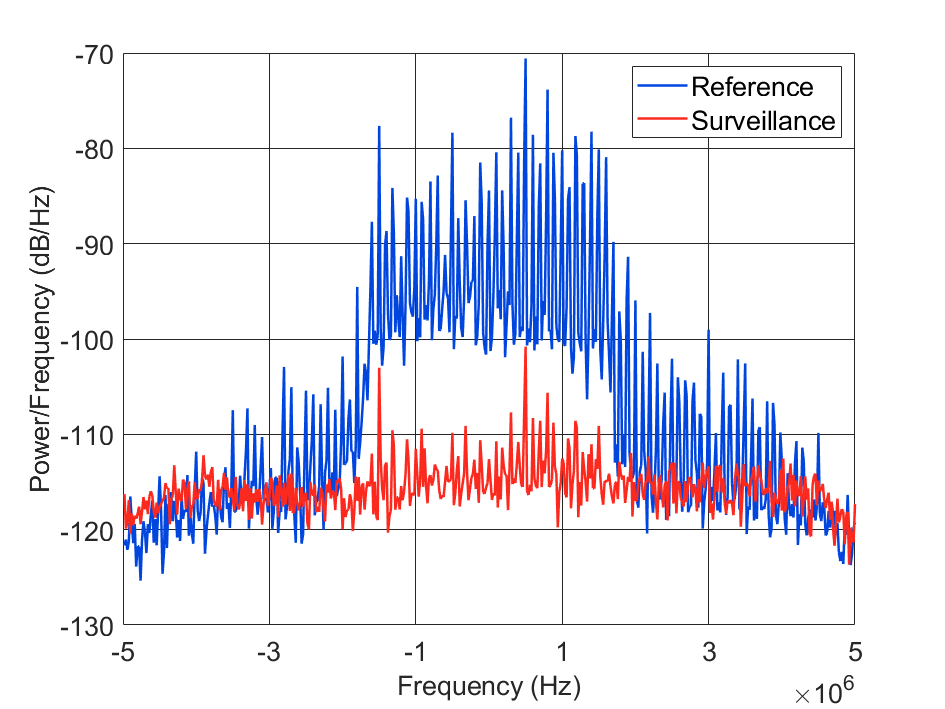}}
	\caption{Power spectral density of received signals in the reference channel and surveillance channel.}
	\label{fig: Power spectral density}
\end{figure}

In this part, the robustness of the passive sensing versus the received SNR is demonstrated. In Fig. \ref{fig: Power spectral density}, the power spectral density (PSD) of both reference channel and surveillance channel in the NLoS scenario is illustrated, where the handwriting is successfully tracked. It can be observed that the average PSD of the surveillance channel is approximately $ 20 $ dB lower than that of the reference channel. The latter SNR level may not be sufficient for the detection of synchronization sequence. In fact, the data obtained from our experiments reveals that the successful rate of packet synchronization in the surveillance channel is less than 10\%.

As a comparison, a high received SNR is required in the existing literature of handwriting tracking via CSI. This is because an accurate estimation of CSI is critical for capturing the phase shift due to the finger motions. As a result, the writing area should be placed close to both the transmitter and receiver in these works. To the best of our knowledge, the handwriting tracking in NLoS scenario has not been demonstrated in the existing literature. 

\subsection{Trajectory Tracking Accuracy}
\begin{figure}[!t] 
	\centering  
	\includegraphics[width=0.6\columnwidth]{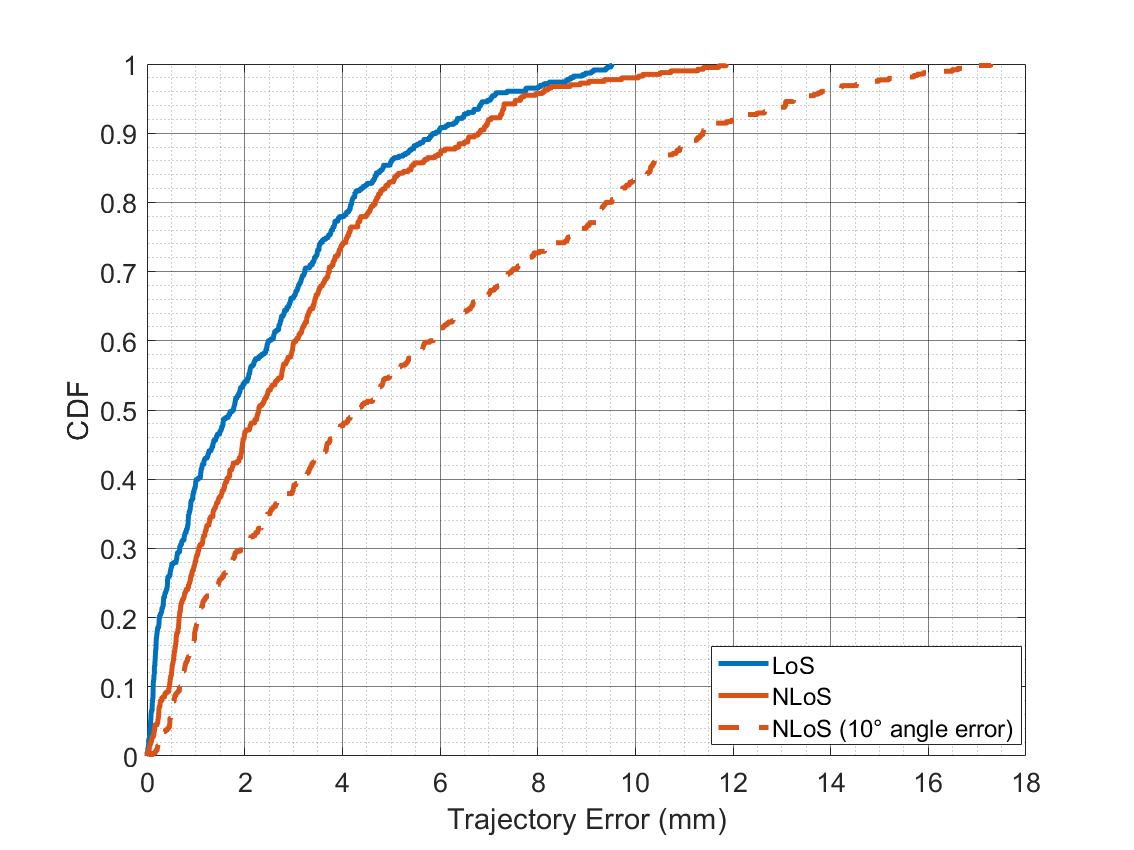}
	\caption{The CDF of trajectory estimation error.}
	\label{fig: CDF}
\end{figure}

In this part, the trajectory tracking errors in all the collected character strokes are compared. Fig. \ref{fig: CDF} illustrates the cumulative distribution functions (CDFs) of these errors in (1) LoS scenario with correct $(x_1,y_1)$, (2) NLoS scenario with correct $(x_1,y_1)$, and (3) NLoS scenario with incorrect $(x_1,y_1)$. It can be observed from the first two CDFs that with correct $(x_1,y_1)$, $ 90 $\% of errors are below $ 6 $ mm and $ 7 $ mm for LoS and NLoS scenarios, respectively. 

Since the coordinates of the initial points $(x_1,y_1)$ are detected via exhaustive beam search, the maximum AoA detection error is less than $10^{\circ}$ (the beam width is $10^{\circ}$). In the third CDF marked by the orange dash curve, we conducted a reconstruction with a $10^{\circ}$ initial AoA measurement error. It can be observed that when the initial AoA measurement error is $10^{\circ}$, the trajectory exhibits a median error of $ 4.5 $ mm, with $ 90 $\% of errors below $ 11.5 $ mm.

\section{CONCLUSION}
In this paper, the cooperative reconstruction of handwriting via an integrated sensing and mmWave communication system is proposed and demonstrated. The system consists of one transmitter (BS) and at least two receivers (UEs), where the handwriting is detected at the two receivers via passive sensing. It is demonstrated that in the LoS scenario, $50$\% of the trajectory errors are below $2$ mm; while in the NLoS scenario, $50$\% of the trajectory errors is below $2.5$ mm even with weak received signal in the surveillance channel.

\bibliographystyle{IEEEtran}
\bibliography{ref}

\begin{thebibliography}{10}
\providecommand{\url}[1]{#1}
\csname url@samestyle\endcsname
\providecommand{\newblock}{\relax}
\providecommand{\bibinfo}[2]{#2}
\providecommand{\BIBentrySTDinterwordspacing}{\spaceskip=0pt\relax}
\providecommand{\BIBentryALTinterwordstretchfactor}{4}
\providecommand{\BIBentryALTinterwordspacing}{\spaceskip=\fontdimen2\font plus
\BIBentryALTinterwordstretchfactor\fontdimen3\font minus
  \fontdimen4\font\relax}
\providecommand{\BIBforeignlanguage}[2]{{%
\expandafter\ifx\csname l@#1\endcsname\relax
\typeout{** WARNING: IEEEtran.bst: No hyphenation pattern has been}%
\typeout{** loaded for the language `#1'. Using the pattern for}%
\typeout{** the default language instead.}%
\else
\language=\csname l@#1\endcsname
\fi
#2}}
\providecommand{\BIBdecl}{\relax}
\BIBdecl

\bibitem{qian2017widar}
K.~Qian, C.~Wu, Z.~Yang, Y.~Liu, and K.~Jamieson, ``{Widar: Decimeter-level
  passive tracking via velocity monitoring with commodity Wi-Fi},'' in
  \emph{Proceedings of the 18th ACM International Symposium on Mobile Ad Hoc
  Networking and Computing}, 2017, pp. 1--10.

\bibitem{IndoTrack}
\BIBentryALTinterwordspacing
X.~Li, D.~Zhang, Q.~Lv, J.~Xiong, S.~Li, Y.~Zhang, and H.~Mei, ``{IndoTrack:
  Device-Free Indoor Human Tracking with Commodity Wi-Fi},'' \emph{Proc. ACM
  Interact. Mob. Wearable Ubiquitous Technol.}, vol.~1, no.~3, sep 2017.
  [Online]. Available: \url{https://doi.org/10.1145/3130940}
\BIBentrySTDinterwordspacing

\bibitem{widraw}
L.~Sun, S.~Sen, D.~Koutsonikolas, and K.-H. Kim, ``Widraw: Enabling hands-free
  drawing in the air on commodity wifi devices,'' in \emph{Proceedings of the
  21st Annual International Conference on Mobile Computing and Networking},
  2015, pp. 77--89.

\bibitem{WiTrace}
L.~Wang, K.~Sun, H.~Dai, A.~X. Liu, and X.~Wang, ``Witrace: Centimeter-level
  passive gesture tracking using wifi signals,'' in \emph{2018 15th Annual IEEE
  International Conference on Sensing, Communication, and Networking
  (SECON)}.\hskip 1em plus 0.5em minus 0.4em\relax IEEE, 2018, pp. 1--9.

\bibitem{CentiTrack}
Z.~Han, Z.~Lu, X.~Wen, W.~Zheng, J.~Zhao, and L.~Guo, ``Centitrack: Towards
  centimeter-level passive gesture tracking with commodity wifi,'' \emph{IEEE
  Internet of Things Journal}, 2023.

\bibitem{PassiveHuman01}
B.~Tan, K.~Woodbridge, and K.~Chetty, ``Awireless passive radar system for
  real-time through-wall movement detection,'' \emph{IEEE Transactions on
  Aerospace and Electronic Systems}, vol.~52, no.~5, pp. 2596--2603, 2016.

\bibitem{PassiveHuman02}
W.~Li, R.~J. Piechocki, K.~Woodbridge, C.~Tang, and K.~Chetty, ``Passive wifi
  radar for human sensing using a stand-alone access point,'' \emph{IEEE
  Transactions on Geoscience and Remote Sensing}, vol.~59, no.~3, pp.
  1986--1998, 2021.

\bibitem{PassiveHuman03}
J.~Li, C.~Yu, Y.~Luo, Y.~Sun, and R.~Wang, ``{Passive motion detection via
  mmWave communication system},'' in \emph{2022 IEEE 95th Vehicular Technology
  Conference:(VTC2022-Spring)}.\hskip 1em plus 0.5em minus 0.4em\relax IEEE,
  2022, pp. 1--6.

\bibitem{mmAlert}
C.~Yu, Y.~Sun, Y.~Luo, and R.~Wang, ``mmalert: mmwave link blockage prediction
  via passive sensing,'' \emph{IEEE Wireless Communications Letters}, pp. 1--1,
  2023.

\bibitem{tan2005passive}
D.~K. Tan, H.~Sun, Y.~Lu, M.~Lesturgie, and H.~L. Chan, ``Passive radar using
  global system for mobile communication signal: theory, implementation and
  measurements,'' \emph{IEE Proceedings-Radar, Sonar and Navigation}, vol. 152,
  no.~3, pp. 116--123, 2005.

\bibitem{Music}
Y.~Sun, J.~Li, T.~Zhang, R.~Wang, X.~Peng, X.~Han, and H.~Tan, ``{An indoor
  environment sensing and localization system via mmWave phased array},''
  \emph{Journal of Communications and Information Networks}, vol.~7, no.~4, pp.
  383--393, 2022.

\bibitem{Multi_Tone}
D.~Vasish, S.~Kumar, and D.~Katabi, ``{Decimeter-Level Localization with a
  Single WiFi Access Point},'' in \emph{13th USENIX Symposium on Networked
  Systems Design and Implementationz(NSDI `16)}, 2016, pp. 165--178.

\end{thebibliography}

\end{document}